\documentclass[useAMS,usenatbib]{mn2e}
\usepackage{times}

\usepackage{graphicx}
\usepackage{url}
\usepackage{listings}
\usepackage{xspace}
\usepackage{booktabs}
\usepackage{tabularx}
\usepackage[multiple]{footmisc}

\newcommand{\phoenix}{\textsc{phoenix}\xspace}
\newcommand{\ldt}{\textsc{LDTk}\xspace}
\newcommand{\ud}{\ensuremath{\mathrm{d}}}

\newcommand{\teff}{\ensuremath{T_\mathrm{Eff}}\xspace}
\newcommand{\logg}{\ensuremath{\log g}\xspace}
\newcommand{\z}{\ensuremath{z}\xspace}

\newcommand{\ldpsetcr}{\texttt{LDPSetCreator}\xspace}
\newcommand{\ldpset}{\texttt{LDPSet}\xspace}

\begin{document}

\title{\ldt: Limb Darkening Toolkit}
\author[H. Parviainen and S. Aigrain]{H. Parviainen\thanks{hannu.parviainen@physics.ox.ac.uk} and S. Aigrain\\Department
of Physics, Denys Wilkinson Building Keble Road, Oxford, OX1 3RH
}

\maketitle

\begin{abstract}
We present a Python package \ldt that automates the calculation of custom stellar limb darkening (LD) profiles and
model-specific limb darkening coefficients (LDC) using the library of \phoenix-generated specific intensity spectra by
\citet{Husser2013}. The aim of the package is to facilitate analyses requiring custom generated limb darkening
profiles, such as the studies of exoplanet transits--especially transmission spectroscopy, where the transit modelling 
is
carried out for custom narrow passbands--eclipsing binaries (EBs), interferometry, and microlensing events. First, \ldt
can be used to compute custom limb darkening profiles with uncertainties propagated from the uncertainties in the
stellar parameter estimates. Second, \ldt can be used to estimate the limb-darkening-model specific coefficients with
uncertainties for the most common limb-darkening models. Third, \ldt can be directly integrated into the log posterior
computation of any pre-existing modelling code with minimal modifications. The last approach can be used to constrain 
the
LD model parameter space directly by the LD profile, allowing for the marginalization over the LD parameter space
without the need to approximate the constraint from the LD profile using a prior.
\end{abstract}


\section{Introduction}
\label{sec:intro}

Stellar limb darkening (LD) is a major source of uncertainty in the characterization of transiting exoplanets
\citep{Csizmadia2013,Espinoza2015}, interferometry-based stellar radius estimation \citep{White2013,Neilson2013a}, and
analysis of microlensing events \citep{An2002}. 

Limb darkening is approximated by a variety of limb darkening models, or \emph{laws} \citep{Mandel2002,Gimenez2006}.
These models aim to reproduce the stellar intensity profile as a function of $\mu$\footnote{Where $\mu = \sqrt{1-z^2} =
\cos \gamma$, $z$ is the normalized distance from the centre of the stellar disk, and $\gamma$ is the foreshortening
angle.} with a relatively small number of parameters, \emph{limb darkening coefficients}.
In the context of exoplanet transit modelling, the early transiting exoplanet studies generally fixed the limb darkening
coefficients to values derived from the model and passband-specific fits to numerical stellar models by
\citet{Claret2000,Claret2004}. However, fixing the coefficients will introduce biases in the parameter estimates if the
models fail to reproduce the true stellar intensity profiles \citep{Csizmadia2013,Espinoza2015}, which has been shown to
be the case in several situations where inferences about the true limb darkening profile have been possible \citep[][but
see also \citealt{Muller2013}]{Fields2003,Claret2008,Claret2009,Howarth2011a}. In addition, even with an accurate model,
fixing the coefficients would lead to underestimated uncertainties since the uncertainties in the stellar parameter
estimates are not propagated into uncertainties in the limb darkening profiles. 

The problems with fixed limb darkening have led to the use of more robust approaches to transit modelling, and currently
the limb darkening is usually either constrained using (more or less informative) priors
based on the tabulated limb darkening coefficients, or using uninformative priors and marginalizing over all the limb
darkening profiles allowed by the observations~\citep{Kipping2013b}. Also, the modern limb darkening tabulations
by \citet{Claret2013,Claret2014}, \citet{Sing2010}, \cite{Neilson2013a,Neilson2013b}, \citet{Magic2014}, and
\citet{Husser2013} are based on more advanced stellar atmosphere models than the early ones, using a spherical model
geometry instead of a plane-parallel one \citep{Claret2012,Claret2014,Neilson2013a,Neilson2013b,Husser2013}, using fully
three-dimensional models geometries \citep{Hayek2012}, and including hydrodynamical simulations to account for the
stellar granulation \citep{Magic2014} .$\!$\footnote{Note that even with reliable stellar models, errors can arise from
the limb darkening model fitting. See~Sect.~2.2 in \citet{Espinoza2015}, especially the need to renormalise the radius
from the spherical models.} 

Allowing the data to speak for itself--using noninformative priors and marginalizing over the limb darkening allowed by 
the data--yields the most reliable parameter estimates, but is not the optimal approach in some situations. For example, 
in transmission spectroscopy, where minute changes in the planetary radius as a function of wavelength are used to make 
inferences about the planet's atmosphere, the uncertainties from unconstrained limb darkening easily drown any variation 
in the radius estimates. Further, the use of pretabulated coefficients is impossible due to the use of narrow 
non-standard passbands, and unaccounted-for small-scale variations in the limb darkening can lead to systematic trends 
and spurious features in the inferred transmission spectrum.

A more robust way to account for limb darkening in any modelling situation is thus to generate limb darkening profiles
from the modelled stellar spectra on a case-by-case basis for each passband in a way that propagates the uncertainties
in the stellar properties to uncertainties in the limb darkening profile, and allows us to set a factor telling how much
we trust on the stellar atmosphere models used to create the spectra. This approach offers a compromise between tightly
constrained and completely unconstrained limb darkening. The limb darkening profiles can be used to either create prior
distributions for the limb darkening model coefficients by estimating the coefficient posteriors; the profiles can be
used to directly constrain the limb darkening in the modelling process, fitting a limb darkening model to the profile
simultaneously with the rest of the analysis; or the profiles can be used directly as is, if the modelling approach
allows for arbitrary stellar limb darkening profiles.

We present a Python package \ldt that automates the calculation of custom stellar limb darkening profiles and
model-specific limb darkening coefficients from the library of \phoenix-generated specific intensity spectra by
\citet{Husser2013}. \ldt propagates the uncertainties in the stellar parameter estimates into the uncertainties in the
limb darkening profile, and offers methods to create multivariate normal priors for the most common limb darkening laws.
The package can also be directly integrated into the modelling code, bypassing the need to pre-calculate the priors.
While the following discussion considers mainly exoplanet transit modelling, the package can be equally used in the
context of eclipsing binary, microlensing, and interferometry studies, or in any work benefiting from custom-built
stellar intensity profiles.

\section{Examples}
\label{sec:examples}
\subsection{Calculation of model coefficients}
\label{sec:examples:coeffs}

We start with an example of limb darkening model coefficient estimation. \ldt offers methods to calculate the maximum
likelihood estimates--as well as the full posterior distributions--for the coefficients of the limb darkening models
listed in Table~\ref{tbl:models}. 

\begin{figure*}
 \centering
 \includegraphics[width=\textwidth]{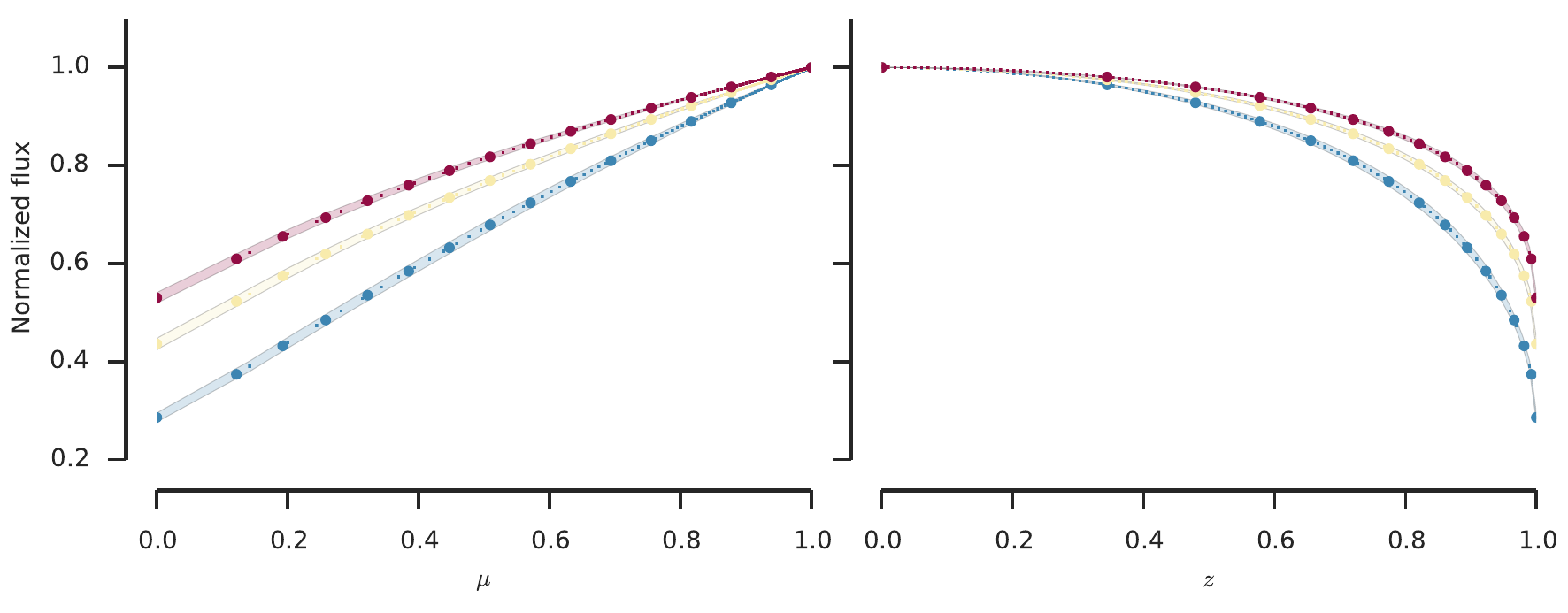}
 \caption{Three limb darkening profiles created by the example code in Sect.~\ref{sec:examples:coeffs} as a function of
$\mu$ (left) and $z$ (right). The large dots show the profile median with the original \citep{Husser2013} sampling, the
small dots show the same resampled to be linear in $z$, and the shaded areas show the $3\sigma$ profile uncertainties.
The fluxes are normalized to unity in the centre of the stellar disk.}
 \label{fig:profiles}
\end{figure*}

\begin{figure*}
 \centering
 \includegraphics[width=\textwidth]{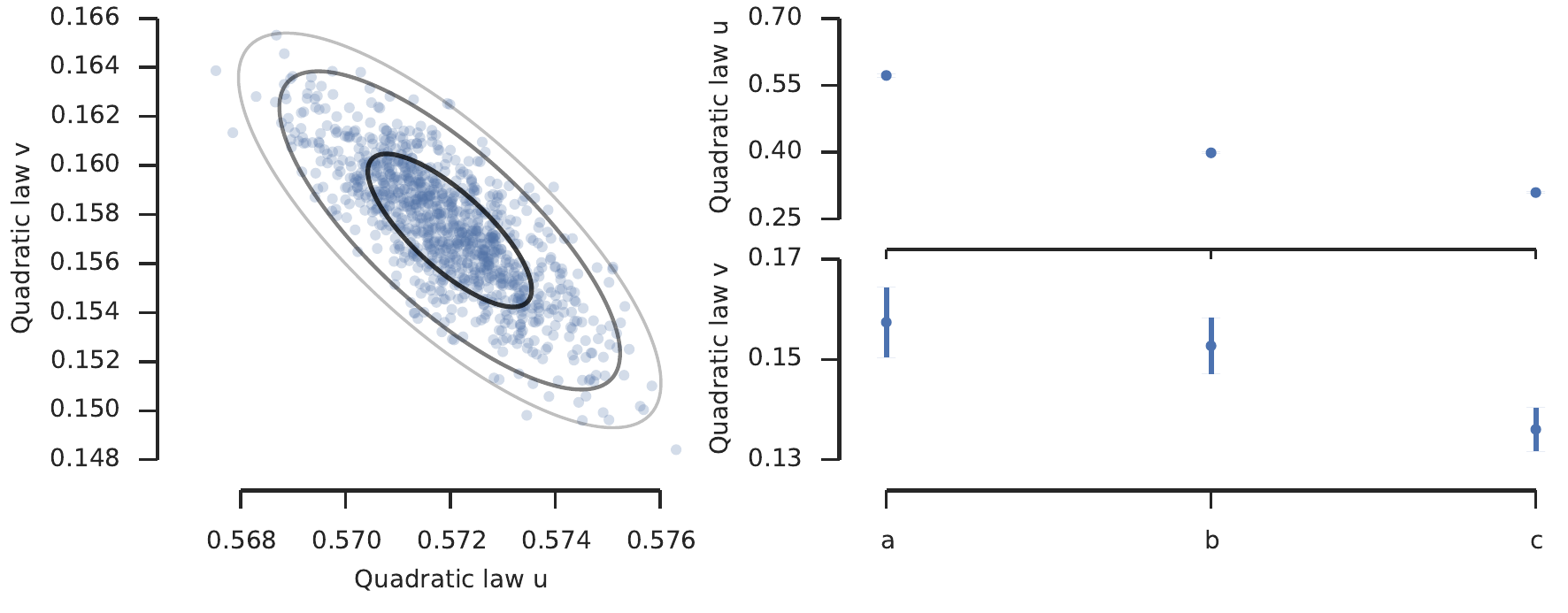}
 \caption{Bivariate normal density approximation for the quadratic model coefficient likelihood (showing 50, 95, and 99
percentile boundaries) and samples from the likelihood calculated using MCMC for the $a$ passband from
Sect.~\ref{sec:examples:coeffs} (left). The coefficient estimates with their uncertainties (99\% central likelihood
intervals) for all three passbands (right). The $u$ errorbars are too small to be visible.}
 \label{fig:joint_uv}
\end{figure*}

\begin{figure}
 \centering
 \includegraphics[width=\columnwidth]{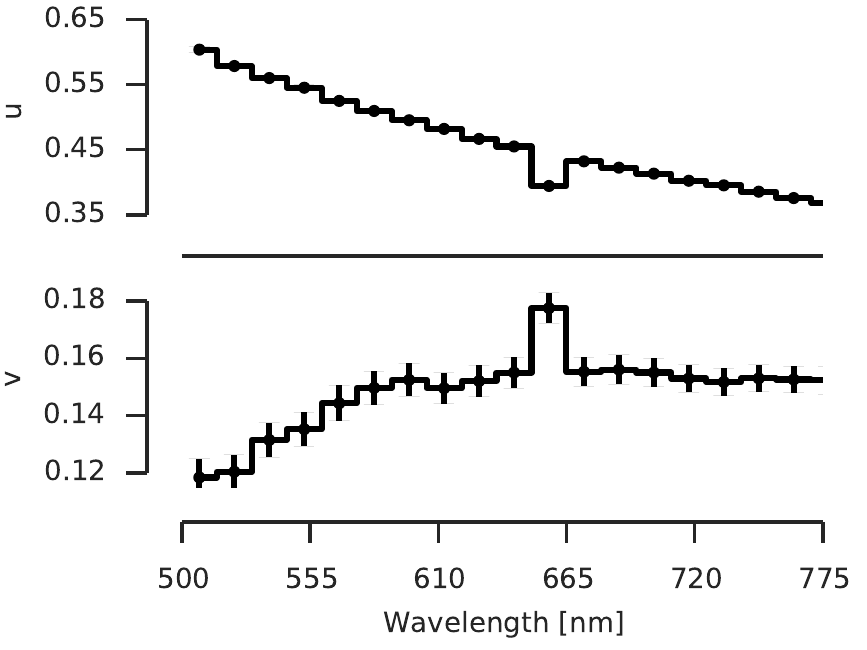}
 \caption{Quadratic limb darkening coefficients for 15~nm wide bins from 500~to~800~nm. Note that the scales are
different for the two figures.}
 \label{fig:narrow_qd_coeffs}
\end{figure}

Limb darkening profiles and quadratic model coefficients for three simple passbands for a star with $\teff =
6400\pm100$~K, $\logg=4.5\pm0.1$, and $\z=0.25\pm0.05$ can be calculated as
\begin{lstlisting}[label=lst:coeffs]
from ldtk import (LDPSetCreator, 
                  BoxcarFilter)
 
filters = [BoxcarFilter('a',450,550),
	   BoxcarFilter('b',650,750),
	   BoxcarFilter('c',850,950)]
	   
sc = LDPSetCreator(filters,
                   teff=[6400,  100], 
                   logg=[4.50, 0.10], 
                      z=[0.25, 0.05])
 
ps = sc.create_profiles(nsamples=500)
qc,qe = ps.coeffs_qd(do_mc=True)
\end{lstlisting}
We first import the \verb|LDPSetCreator| class that will create a set of limb darkening profiles for a given set of
filters. Next, we define the filters, here using simple boxcar filters that have zero transmission outside the given
minimum and maximum wavelength range. We continue by creating an instance of \verb|LDPSetCreator|, initialising it with
the filter set and stellar parameter estimates. \verb|LDPSetCreator| then creates a list of needed spectrum files,
checks them against the local cache, and downloads the uncached files from the \citet{Husser2013} FTP library. After
this, \verb|sc.create_profiles| is used to create the limb darkening profiles for each filter, shown in
Fig.~\ref{fig:profiles}, contained in an instance of \verb|LDPSet| class. The \verb|LDPSet| class is finally used to 
estimate the quadratic limb darkening model coefficients \verb|qc| and their uncertainty estimates \verb|qe|,
illustrated in Fig.~\ref{fig:joint_uv}. 
Figure \ref{fig:narrow_qd_coeffs} shows the quadratic model coefficients for the same star, but calculated for 19 narrow
(15~nm wide) passbands from 500~to~800~nm.

\subsection{Integration into transit modelling}
\label{sec:example:integration}
Rather than calculating the limb darkening coefficients prior to modelling, \ldt can be integrated directly
into the code carrying out the log posterior evaluation. This can be done in two ways: (1) \ldt can be used to
evaluate the log likelihood for the limb darkening model given model coefficients during the posterior sampling; (2)
\ldt can be used to create a multivariate normal prior automatically in the initialisation before the sampling.

\subsubsection{Log likelihood evaluation}
\label{sec:example:integration:lnlike_evaluation}
The log likelihood for an LD profile can be calculated as
\begin{lstlisting}[label=lst:basic]
lnl_ld = ps.lnlike_xx(cf)
\end{lstlisting}
where \verb|cf| is a coefficient array, and \verb|xx| is the model abbreviation from Table~\ref{tbl:models}. The log
likelihood call can be integrated into the log posterior computation directly
\begin{lstlisting}
class LPFunction(object):
  def __init__(self, ldfile=None, ...):
    << OMITTED                  >>
    << set up the data, models, >>
    << parameterisation, etc.   >>
    
    if not exists(ldfile):
      sc = LDPSetCreator(...)             
      self.ps = sc.create_profiles()
      self.ps.save(ldfile)
    else:
      self.ps = load_ps(ldfile)   
    self.lnlike_ld = self.ps.lnlike_qd
   
  def lnprior(self, pv):
    << calculate the prior >>
    
  def lnlike(self, pv):
    << calculate the log likelihood >>
    << for the main dataset(s)      >>
    
  def __call__(self, pv):
    return (self.lnprior(pv) +
            self.lnlike(pv) +
            self.lnlike_ld(pv[ldsl]))
\end{lstlisting}
where \verb|LPFunction| is a callable class that encapsulates the log posterior computation. The limb darkening
profiles are created during the first initialisation, and saved to a file so that we do not need to regenerate them
every time. After the initialisation, all that is required is to add a call to calculate the log likelihood to the
calculation of the log posterior. The example implements this in the \verb|__call__| method, where \verb|pv| is the
model parameter vector, and \texttt{ldsl} is a slice selecting the limb darkening coefficients from the parameter
vector.

\subsubsection{Multivariate normal prior generation}
\label{sec:example:integration:prior_generation}

The second approach for the direct integration of \ldt is to use it to construct a multivariate normal prior in the
initialisation
\begin{lstlisting}[label=lst:example:prior_generation]
<< class initialization >>
self.lnp_ld = self.ps.create_prior_xx()

<< log posterior evaluation >>
return (self.lnprior(pv) +
	self.lnp_ld(pv[lds]) +
	self.lnlike_lc(pv))
\end{lstlisting}
where \verb|lnp_ld| is now a function that returns the joint log density of multivariate normal priors estimated from
the profiles.

\section{Implementation}
\label{sec:implementation}
\subsection{Overview}
\label{sec:implementation:overview}

The package aims to be simple to install and integrate into pre-existing Python-based modelling code, and
depends only on standard Python modules, NumPy, SciPy, and PyFITS. The code with IPython-notebook-based examples is
freely available from
\begin{quote}
 \url{https://github.com/hpparvi/ldtk}
\end{quote} 
The repository contains also examples of the package's basic usage, ways to integrate the package into a transit
light curve analysis, and notebooks detailing the implementation of the code.

\subsection{Profile creation}
\label{sec:implementation:profile_creation}

The limb darkening profiles are created using the \phoenix-calculated specific intensity spectra by \citet{Husser2013}.
The spectra are stored in an FTP server as FITS files--a single file for a single set of stellar parameters--where
each file contains a model spectrum spanning the wavelength range from 50~to~2600~nm with a 0.1~nm resolution for 78
values of $\mu$. Each specific intensity file is $\sim$15~MB, and the whole library has a size of hundreds of GBs.
However, downloading the whole library is unnecessary, and only a small subset of spectra are required for any
given star.

The profile creation starts with the calculation of the necessary spectrum files, continues with a check testing which
of the required files are already in a local cache directory, after which the missing files are downloaded from the
library FTP server. The \verb|LDPSetCreator| class carries out these tasks during its initialisation, including all
stellar spectra inside a $2n\sigma$-wide cube on the three stellar parameters, where $\sigma$ are the stellar
parameter uncertainties, and $n$ is a freely set factor defaulting to three.

Next, limb darkening profiles are calculated for each filter and stellar parameter set as 
\begin{equation}
 I(\mu) = \int Q(\lambda) F(\lambda) I(\mu,\lambda) \;\ud\lambda,
\end{equation}
where $Q$ is the detector quantum efficiency, $F$ is the filter transmission, and $I(\mu,\lambda)$ is the stellar
spectrum.
The profiles are renormalized in $z$ \citep[see][Sect.~2.2]{Espinoza2015}, and then used to construct a
 ($\teff,\logg,\z$) interpolation grid for each filter.

The final limb darkening profiles are calculated using Monte Carlo sampling. The \verb|create_profiles| method generates
$N$ stellar parameter samples from a multivariate normal distribution with the means and variances defined by the
stellar parameter estimates and their uncertainties (we assume symmetrical normal uncertainty distributions), and
calculates the limb darkening profiles for each sample. The final profile for each filter is defined by the mean of the
profile set, and the profile uncertainty by its standard deviation.

\subsection{Limb darkening coefficient estimation}
\label{sec:implementation:ldc_estimation}

\begin{table}
\caption{Implemented limb darkening models, their abbreviations and definitions.}
\label{tbl:models}
\begin{center}
\begin{tabularx}{\columnwidth}{@{\extracolsep\fill}lll}
\toprule
Name     & Abbr. & Model \\
\midrule
Linear    & \texttt{ln} & $1 - C(1-\mu)$ \\
Quadratic & \texttt{qd} & $1 - C_u(1-\mu) - C_v(1-\mu)^2$\\
Nonlinear & \texttt{nl} & $1 - \sum_{n=1}^4 c_n (1-\mu^{n/2})$ \\
General   & \texttt{ge} & $1 - \sum_{n=1}^N c_n (1-\mu^n)$ \\
\bottomrule
\end{tabularx}
\end{center}
\end{table} 

The \verb|LDPSet| class offers methods of form \verb|coeffs_xx|, where \verb|xx| is a limb darkening model abbreviation
from Table~\ref{tbl:models}, to estimate the limb darkening model coefficients with their uncertainties. For example,
the quadratic model coefficients can be obtained as
\begin{lstlisting}
qc,qe = ps.coeffs_qd()
\end{lstlisting}
where \verb|qc| contains the coefficient estimates and \verb|qe| their uncertainties for each filter. By default, the
estimates correspond to the maximum likelihood estimates, and the uncertainties to the 68\% central likelihood intervals
assuming multivariate normal likelihood with zero correlation between the coefficients. 
That is, we estimate the likelihood uncertainty as
\begin{equation}
 \sigma = \sqrt{-\frac{\ud^2 \ln P(x_0)}{\ud^2x}},
\end{equation}
where $P$ is the likelihood density, $x_0$ is the maximum likelihood estimate for $x$, $\sigma$ its standard deviation,
and the derivative is calculated numerically.

More robust uncertainty estimates can be obtained using Markov Chain Monte Carlo (MCMC) sampling
\begin{lstlisting}
qc,qe = ps.coeffs_qd(do_mc=True)
\end{lstlisting}
which also allows for the estimation of the covariance matrices
\begin{lstlisting}
qc,qm = ps.coeffs_qd(do_mc=True,
                     return_cm=True)
\end{lstlisting}
which can be useful when the estimates are used as priors in the modelling process. 

The methods also allow the MCMC parameter estimation to be fine tuned by setting the number of iterations, the chain
thinning factor, and the number of burn-in iterations, and the MCMC chain is stored inside the \verb|LDPSet| instance to
allow for a more detailed convergence analysis.

\subsection{Log likelihood evaluation}
\label{sec:implementation:lnlike_evaluation}
The \verb|LDPSet| class offers the methods \verb|lnlike_xx|, where \verb|xx| is again the model abbreviation, for the 
evaluation of the model likelihoods given a set of model coefficients. The joint likelihood for a quadratic model can be
calculated as
\begin{lstlisting}
ll = ps.lnlike_qd([[0.35,0.15],
                   [0.23,0.11],
                   [0.18,0.10]])
\end{lstlisting}
where we evaluate the log likelihood with two pairs of coefficients, one pair for each filter in the set considered in
the first example.

The likelihoods are calculated assuming that the profile values for a given $\mu$ are normally distributed, which leads
to the usual log likelihood equation
\begin{equation}
\label{eq:logl}
 \log L = -\frac{N\log{2\pi}}{2} - \sum_{i=1}^N \log \epsilon\sigma_i - \sum_{i=1}^N
\frac{(P_i-M_i)^2}{2\epsilon^2\sigma_i^2},
\end{equation}
where $N$ is the number of datapoints, $\sigma_i$ are the profile uncertainties, $P_i$ are the profile values, $M_i$ are
the model values, and $\epsilon$ is a factor that can be used to increase the uncertainties based on how much we trust
the stellar spectrum models.

\subsection{Uncertainty multiplier}
\label{sec:implementation:uncertainty_multiplier}
The uncertainty multiplier $\epsilon$  is a subjective factor that defines how strongly the limb darkening profile (or
the prior created from it) constrains the final analysis (that is, how much we trust the stellar atmosphere models
used to create the profiles.) The uncertainty multiplier is applied to log likelihood calculations and model coefficient
uncertainty estimation as described in Eq.~\ref{eq:logl}.


\subsection{Resampling}
\label{sec:implementation:resampling}
The sampling in $\mu$ used by \citet{Husser2013} focuses on the detailed sampling of the stellar edge, while the inner
parts of the stellar disk are sampled sparsely. While sampling like this is a good choice for representing the stellar
intensity profile, it may not be optimal for likelihood evaluation and model fitting (at least without sample
weighting), since the fit will be dominated by the very edge. A linear sampling in~$z$, for example, may be a better
approach. \ldpset includes a method to resample the profiles to any given vector of $\mu$ or $z$ values,
\verb|resample(mu=mu_array,z=None)|, and also two utility methods \verb|resample_linear_mu(nsamples)| and
\verb|resample_linear_z(nsamples)| to resample the profiles linearly in $\mu$ or $z$.

\subsection{Filters}
\label{sec:implementation:filters}
A filter defines a single passband, and can be either a function or a callable class that returns an array of
transmission values between zero and one given an array of wavelengths in nm. The package comes with some utility
filters, such as the  \verb|BoxcarFilter| used in the first example, and the \verb|TabulatedFilter|, which allows for
the use of tabulated transmission values listed as a function of wavelength.

\subsection{Quantum efficiency}
\label{sec:implementation:qe}
Any function that evaluates the quantum efficiency as a function of wavelength (similar to a filter) can be used to
define the instrumental quantum efficiency. \ldpsetcr can be given a detector quantum efficiency curve in the
initialisation, and a \verb|TabulatedQE| class can be used the same way as \verb|TabulatedFilter|.

\section{Discussion}
\label{sec:discussion}
While our ability to model stellar limb darkening is improving due to the advances in the stellar atmosphere models, we 
still need to develop the ways we integrate the information from these models into our analyses. Current stellar 
spectrum libraries allow us to calculate limb darkening profiles for custom passbands while propagating the 
uncertainties in the stellar parameter estimates into the limb darkening profile uncertainties, and we should take 
advantage of this capability. The marginalization over all the limb darkening profiles allowed by the observations 
yields the most robust (or, at least, conservative) parameter estimates, but the use of information from the stellar 
models is also justified in many situations. Transmission spectroscopy, future planet-characterization space missions 
such as TESS~\citep{Ricker2014} and Plato \citep{Rauer2014}, stellar interferometry, and microlensing studies will all 
benefit from more advanced ways to integrate the results from stellar atmosphere modelling into the analysis 
process.

\begin{figure*}
 \centering
 \includegraphics[width=\textwidth]{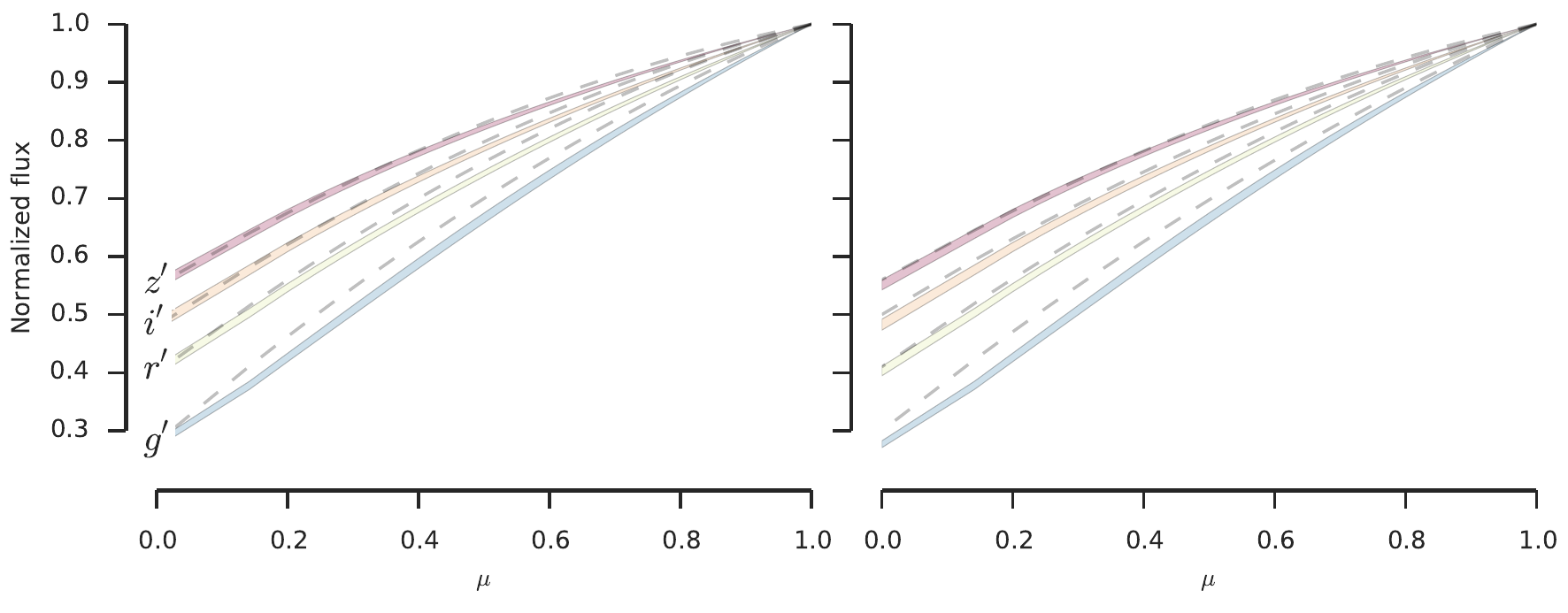}
 \caption{Limb darkening profiles for a $\teff=6500 \pm 50~K$, $\logg=4.5\pm0.1$, and $\z=0.0\pm0.05$ star observed in 
$g'\!$, $r'\!$, $i'\!$, and $z'\!$ filters (coloured areas), and the corresponding quadratic model fits from 
\citet{Claret2011} using either the least-squares (left) or flux conservation method (right).}
 \label{fig:claret_profiles}
\end{figure*}

The \ldt-calculated profiles and model coefficients cannot be directly compared with the previous tabulations by 
\citet{Claret2013} or others due to the differences in handling of the geometry and sampling. Especially the 
redefinition of the stellar edge \citep[see][Sect. 2.2]{Espinoza2015} and the use of uniform sampling in $z$-space will 
affect the model coefficient estimates (however, these both can be changed by the user). 
Figure~\ref{fig:claret_profiles} shows \ldt-created profiles for a $\teff=6500 \pm 50~K$, $\logg=4.5\pm0.1$, and 
$\z=0.0\pm0.05$ star observed in $g'\!$, $r'\!$, $i'\!$, and $z'\!$ filters with the corresponding quadratic models 
using the tabulations by \citet{Claret2013}. The \citeauthor{Claret2013} fits correspond to spherical \textsc{phoenix} 
models where the intensities corresponding to $\mu < 0.1$ have been removed from the fit. The \citet{Claret2013} models 
agree with the \ldt results close to the stellar limb, but deviate significantly for a large span of $\mu$ (and thus for 
most part of the stellar disk) in the blue passbands. 

\section{Conclusions}
\label{sec:conclusions}
We have presented a Python package \ldt that aims to facilitate the inclusion of information from the \citet{Husser2013}
stellar atmosphere model spectrum library into astrophysical modelling problems. \ldt offers a modelling approach where
the information from the stellar atmosphere models can be used to constrain the stellar limb darkening softly, the
strength of the constraint being defined by the researcher's (subjective) trust to the stellar models. The package can
be used to calculate limb darkening coefficients and their uncertainties for freely defined passbands prior to the
modelling, or it can be directly integrated into the modelling code. \ldt can also be used to generate limb darkening
profile samples directly, bypassing the need to use simple limb darkening models if the modelling approach allows for
the use of arbitrary limb darkening profiles \citep[such as the \textsc{batman} transit modelling
code,][]{Kreidberg2015}. This can be useful when the features that are not well captured by the simple limb darkening
models, such as the very edge of the star, are important.

\bibliographystyle{mn2e} 
\bibliography{ldtk}

\end{document}